\documentclass{aastex63}
\pdfoutput=1
\usepackage[T1]{fontenc} 
\usepackage[latin1]{inputenc}
\usepackage{graphicx}	% Including figure files
\usepackage{amsmath}	% Advanced maths commands
\usepackage{amssymb}	% Extra maths symbols \usepackage{natbib}
\usepackage{url} 
\usepackage{appendix}
\usepackage{longtable}
\usepackage{aas_macros} 
\usepackage{lineno}
\newcommand{\cdthree}{2020~CD$_{3}$} % 2020 CD3
\newcommand{\RH}{2006~RH$_{120}$} % 2006 RH120

\newcommand{\rcom}[1]{#1}

\begin{document}
%\maketitle

\title{Establishing Earth's minimoon population through characterization
of asteroid 2020~CD$_3$}

\author{Grigori Fedorets} \affiliation{Astrophysics Research Centre,
School of Mathematics and Physics, Queen's University Belfast, Belfast
BT7 1NN, UK}

%$^{1}$\thanks{Corresponding author. E-mail: fedorets@iki.fi},
\author{Marco Micheli}
\affiliation{ESA NEO Coordination
Centre, Largo Galileo Galilei, 1, 00044 Frascati (RM), Italy}
\affiliation{INAF - Osservatorio Astronomico di Roma, Via Frascati, 33,
00040 Monte Porzio Catone (RM), Italy}

\author{Robert Jedicke} \affiliation{University of Hawai`i,
Institute for Astronomy, 2680 Woodlawn Drive, Honolulu, Hawai`i, 96822,
USA} 
\author{Shantanu P. Naidu} \affiliation{Jet Propulsion
Laboratory, California Institute of Technology, Pasadena, CA, 91109,
USA}

\author{Davide Farnocchia} \affiliation{Jet Propulsion
Laboratory, California Institute of Technology, Pasadena, CA, 91109,
USA}

\author{Mikael Granvik} 
\affiliation{Department of Physics,
P.O. Box 64, 00014 University of Helsinki, Finland}
\affiliation{Asteroid Engineering Laboratory, Onboard Space Systems,
Lule\aa{} University of Technology, Box 848, 98128 Kiruna, Sweden}

\author{Nicholas Moskovitz} \affiliation{Lowell Observatory,
1400 W Mars Hill Road, Flagstaff, AZ 86001, USA}

\author{Megan E. Schwamb} \affiliation{Astrophysics Research
Centre, School of Mathematics and Physics, Queen's University Belfast,
Belfast BT7 1NN, UK} 

\author{Robert Weryk} \affiliation{University of Hawai`i,
Institute for Astronomy, 2680 Woodlawn Drive, Honolulu, Hawai`i, 96822,
USA}

\author{Kacper Wierzcho\'{s}} \affiliation{The University of
Arizona, Lunar and Planetary Laboratory, 1629 E. University
Blvd.,Tucson, AZ 85721, USA}

\author{Eric Christensen} \affiliation{The University of
Arizona, Lunar and Planetary Laboratory, 1629 E. University
Blvd.,Tucson, AZ 85721, USA}

\author{Theodore Pruyne} \affiliation{The University of Arizona,
Lunar and Planetary Laboratory, 1629 E. University Blvd.,Tucson, AZ
85721, USA}

\author{William F. Bottke} \affiliation{Department of Space
Studies, Southwest Research Institute, 1050 Walnut Street, Suite 300,
Boulder, CO 80302, USA}

\author{Quanzhi Ye} \affiliation{Department of Astronomy,
University of Maryland, College Park, MD 20742, USA}

\author{Richard Wainscoat} \affiliation{University of Hawai`i,
Institute for Astronomy, 2680 Woodlawn Drive, Honolulu, Hawai`i, 96822,
USA}

\author{Maxime Devog\`ele} \affiliation{Lowell Observatory, 1400
W Mars Hill Road, Flagstaff, AZ 86001, USA}

\author{Laura E. Buchanan} \affiliation{Astrophysics Research
Centre, School of Mathematics and Physics, Queen's University Belfast,
Belfast BT7 1NN, UK}

\author{Anlaug Amanda Djupvik} \affiliation{Nordic Optical
Telescope, Apartado 474, E-38700 Santa Cruz de La Palma, Santa Cruz de
Tenerife, Spain}

\author{Daniel M. Faes} \affiliation{Gemini Observatory/NSF's
NOIRLab, 670 N. A'ohoku Place, Hilo, Hawai'i, 96720, USA}

\author{Dora F\"{o}hring} \affiliation{University of Hawai`i,
Institute for Astronomy, 2680 Woodlawn Drive, Honolulu, Hawai`i, 96822,
USA}

\author{Joel Roediger} \affiliation{National Research Council
of Canada, Herzberg Astronomy and Astrophysics Research Centre, 5071
West Saanich Road, Victoria, BC V9E 2E7, Canada}

\author{Tom Seccull} \affiliation{Gemini Observatory/NSF's
NOIRLab, 670 N. A'ohoku Place, Hilo, Hawai'i, 96720, USA}

\author{Adam B. Smith} \affiliation{Gemini Observatory/NSF's
NOIRLab, 670 N. A'ohoku Place, Hilo, Hawai'i, 96720, USA}

\correspondingauthor{Grigori Fedorets} \email{fedorets@iki.fi}

\begin{abstract} We report on our detailed characterisation of Earth's
second known  temporary natural satellite, or minimoon, asteroid
\cdthree{}. An artificial origin can be ruled out based on its
\rcom{area-to-mass ratio and broad-band photometry}, which suggest that
it is a silicate asteroid belonging to the S or V complex in asteroid
taxonomy. The discovery of \cdthree{} allows for the first time a
comparison between known minimoons and theoretical models of their
expected physical and dynamical properties. The estimated diameter of
$1.2^{+0.4}_{-0.2}$ meters and geocentric capture approximately a decade
after the first known minimoon, \RH, are in agreement with theoretical
predictions. The capture duration of \cdthree{} of at least 2.7 years is
unexpectedly long compared to the simulation average, but it is in
agreement with simulated minimoons that have close lunar encounters,
providing additional support for the orbital models. \cdthree{}'s
atypical rotation period, significantly longer than theoretical
predictions, suggests that our understanding of meter-scale asteroids
needs revision. More discoveries and a detailed characterisation of the
population can be expected with the forthcoming Vera C. Rubin
Observatory Legacy Survey of Space and Time (LSST). \end{abstract}

\section{Introduction}

Asteroids and comets can be temporarily captured by planets as natural
satellites. Theoretical models \citep{granvik2012,fedorets2017} predict
that the Earth is also surrounded by a cloud of such temporarily
captured asteroids, colloquially called minimoons. The largest minimoon
captured at any given time is one meter in diameter, while larger bodies
are captured less frequently.

Minimoons possess a number of attributes which make them objects of
particular interest. As they spend an extended amount of time in the
vicinity of the Earth, they can provide several windows of opportunity
to obtain observations of the little studied population of meter-class
asteroids. The systematic discovery and population statistics of meter
to decameter-class minimoons, a sub-population of the near-Earth objects
(NEO), could resolve existing disagreements between extrapolations of
different NEO size-frequency distribution models to this size range -- 
i.e. those based, on telescopic observations \citep[e.g.,
][]{rabinowitz2000,harris2015,granvik2016a,tricarico2017} and those
based on on bolide data \citep{brown2002,brown2013}. Also, due to their
relatively long capture duration, accessibility, and small size,
minimoons are viable targets for taking the first practical steps in the
emerging field of asteroid \textit{in situ} resource utilisation
\citep{granvik2013,jedicke2018}. So far, the primary obstacle for
organizing their study  has been a lack of observational evidence
supporting the existence of a minimoon population to the extent
predicted by the models.

Until 2020, only one minimoon \citep[\RH
;][]{bressi2008,kwiatkowski2009} had been discovered.  The second known
minimoon, \cdthree{}, was discovered on 2020 February 15.51 UT by the
Catalina Sky Survey \citep[CSS;][]{christensen2018} 1.5 m telescope on
Mt. Lemmon \citep{mpec2020-d104}. One day later, an alert automatically
sent out by the NASA Jet Propulsion Laboratory's Scout system
\citep{farnocchia2015,farnocchia2016} announced that it was likely
temporarily captured in the Earth-Moon system. Discovering an object on
a geocentric orbit always raises suspicion of an artificial origin, but
during the 2.5~weeks following its discovery \cdthree{} could not be
linked to any known artificial object nor could a natural origin be
ruled out. On 2020 February 26 the Minor Planet Center (MPC) therefore
added \cdthree{} to the catalogue of asteroids as a temporarily-captured
object with a request for further follow-up observations to establish
its nature \citep{mpec2020-d104}.

\begin{figure}[h] \centering
\includegraphics[width=0.6\textwidth]{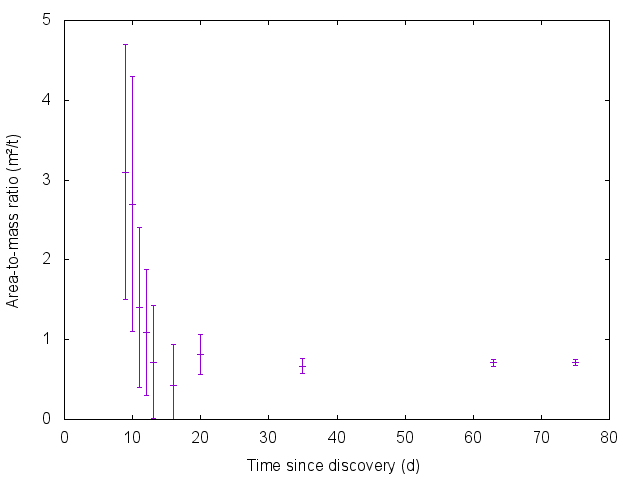} \caption{The
evolution of the detection of the zero-albedo area-to-mass ratio with
1$\sigma$ errorbars as the function of the length of the observational
arc.} \label{fig:amr_evolution} \end{figure}

The nominal solution for the area-to-mass ratio of \cdthree{} ---
calculated from the solar radiation pressure signature on the orbital
solution, and a diagnostic quantity for distinguishing between natural
and artificial objects \citep{jedicke2018} --- decreased during the two
weeks after discovery (Fig.~\ref{fig:amr_evolution}), indicating that it
might be a natural object. To characterize the potential minimoon, we
obtained high-precision astrometric follow-up observations in
February-May 2020 with the Nordic Optical Telescope (NOT),
Canada-France-Hawai'i Telescope (CFHT), Lowell Discovery Telescope
(LDT), University of Hawai'i 2.2 m telescope (UH2.2) and the Calar Alto
Schmidt telescope; broad-band photometric observations from Gemini
North; and rotational lightcurve observations with LDT. In addition, a
search for pre-discovery detections with the Pan-STARRS surveys
\citep{chambers2016panstarrs1}, Zwicky Transient Facility
\citep{Bellm2019,Masci2019}, Catalina Sky Survey
\citep{christensen2018}, and Chinese Near-Earth Object Survey Telescope
\citep{Zhao2007} was performed.

In this work, we provide a detailed characterization of the physical
properties and orbital evolution of \cdthree{}. We also discuss its
detectability and assess the possibility of its lunar origin. We
describe the observations and precovery attempts in detail in
Sect.~\ref{s:acquisition}; outline the data reduction and methods for
physical characterization and orbit computation in
Sect.~\ref{s:methods}; present the results and discuss the implications
in Sect.~\ref{s:rd}; and offer our conclusions in
Sect.~\ref{s:conclusions}.

\section{Data acquisition} \label{s:acquisition}

\subsection{Instruments and observations}

An overview of all instruments used in this analysis is provided in
\rcom{Appendix} Table~\ref{tab:telescope}. \cdthree{} was discovered on
2020 February 15.51 UT by the Catalina Sky Survey
\citep[CSS;][]{christensen2018} 1.5~m telescope on Mt. Lemmon (MPC
observatory code G96). Upon discovery, the object was favorably placed
near the ecliptic plane $\sim$45$^{\circ}$ east of opposition. The
discovery image sequence consisted of four 30 s exposures, with $\sim7$
minutes separation between each successive image, that were inspected
soon after the final image by two observers and submitted to the MPC as
a new NEO candidate. After the object was placed on the MPC's NEO
Confirmation Page, additional same-night follow-up observations were
performed with the \rcom{same telescope that was used to discover
\cdthree{}.}

Multiple broadband photometric imaging was performed on 2020 February 24
with the  8.1~m Frederick C. Gillett Gemini North Telescope located on
Maunakea, Hawai`i, USA. The Gemini Multi-Object Spectrograph
\citep[GMOS;][]{2004PASP..116..425H}  observations consisted of three
$r'g'i'$ sequences with the G0301, G0303, and G0302 filters, taken in
photometric conditions with Image Quality 85 (1.05$^{\prime\prime}$ full
zenith corrected seeing) or better seeing. The telescope was tracked
non-sidereally at \cdthree{}'s rate of motion, thereby maintaining its
stellar point spread function (PSF) for photometry but elongating the
reference field stars. We also obtained sidereally tracked images in the
three filters immediately before and after the non-sidereal tracking of
\cdthree{} in order to perform absolute photometry.

To obtain \cdthree{}'s photometric lightcurve  we employed the Large
Monolithic Imager (LMI) on the 4.3~m Lowell Discovery Telescope (LDT,
G37) for approximately 1 hour on 2020 February 27 UTC.  Exposures were
taken with 30~s integrations using a broad band $VR$ filter that
provides high throughput between approximately 500 and 700 nm. LMI was
binned $3 \times 3$ for an effective plate scale of 0.36"/pixel and the
telescope was tracked at the non-sidereal rates of the target. On
several later occasions we used LDT/LMI to obtain astrometry with a
similar technique but without any filters.

The 3.6~m Canada France Hawaii Telescope (CFHT, 568) on Maunakea,
Hawai`i, USA, was used to obtain astrometry using non-sidereal tracking
with exposures of up to 120~s in {\it gri}-band MegaCam images with no
pixel binning.  MegaCam has $0.187^{\prime\prime}$ pixels allowing for
precise astrometric measurements under good seeing conditions.

Astrometric observations were also made with the Alhambra Faint Object
Spectrograph and Camera (ALFOSC) at the 2.5~m Nordic Optical Telescope
(NOT, Z23) at the Roque de los Muchachos Observatory, La Palma, Canary
islands, Spain. The exposures were tracked non-sidereally on \cdthree{}.
Each image's exposure time was set equal to the time it would take for
\cdthree{} to move at most one stellar FWHM on the sky. Most of the
\cdthree{} detections had $S/N\ge15$ but the last observations reached
only $S/N\sim5$ as the target reached the detection threshold. The
observations were performed without any filters with $4 \times 4$ pixel
binning.

The University of Hawaii 2.2~m (UH2.2, 568) telescope was used for
astrometric observations with non-sidereal tracking at \cdthree{}'s
apparent rates of motion in unfiltered 300~s exposures. 
Additional astrometric observations were extracted from dedicated early
observations obtained with the Calar Alto Schmidt telescope (Z84) in
Spain. The detections were obtained from a set of short sidereally
tracked frames, stacked with respect to the known motion of the object.

\subsection{Search for pre-discovery detections} % -- ROBERT W.

The image archives for several survey telescopes were searched for
pre-discovery observations of \cdthree{} by generating an ephemeris for
each exposure and visually examining any potential matches. The 1.8~m
Pan-STARRS1 telescope (F51) has an extensive archive dating back to 2010
\citep{chambers2016panstarrs1} and is sensitive to $V\sim23$ but no
detections were found. The Zwicky Transient Facility (ZTF, I41) is an
ongoing wide-field optical survey using the 1.2~m Palomar Oschin Schmidt
telescope and has been in operation since 2018 \citep{Bellm2019,
Graham2019, Masci2019}.   No detections were found in its Data Release 3
(DR3) archive that extends from March 2018 to the end of December 2019.

The Chinese NEO Survey Telescope \citep[CNEOST, D29;][]{Zhao2007} is a
1.0~m Schmidt telescope at Xuyi, Jiangsu, China, equipped with a
$3^\circ \times 3^\circ$ camera.  We searched images taken between
January 2018 and May 2019 (when the telescope went offline for hardware
upgrades) but did not find any matching fields. Lastly, we checked all
the telescopes used by the Catalina Sky Survey for pre-discovery
opportunities and found only two suitable fields imaged by the Mt.
Lemmon telescope (G96) on 2019 November 9 and 2019 January 24, close to
times when the object was expected to be at perigee and therefore
relatively bright. Significant trailing losses, the spreading of the
light from the target over many pixels due to its motion during an
exposure, combined with non-optimal sky conditions, prevented a
detection in both images. In summary, the signals in the possible images
were mostly smeared by trailing losses, and no detections were found
from any of the mentioned surveys.

\section{Data reduction and calculations} \label{s:methods}

\subsection{Astrometric data reduction}

Due to the different observing strategies and capabilities of each
instrument/telescope combination, each image set was astrometrically
analyzed with different techniques. In some cases, a direct measurement
on individual frames was possible by fitting \cdthree's detection to a
stellar PSF or trail. In other cases, especially later in the
apparition, we stacked multiple frames at \cdthree's  (often rapidly
changing) rates of motion to achieve sufficient SNR for a measurable
detection. We carefully estimated our formal astrometric uncertainty
taking into account contributions from the object's SNR (often
dominant), but also from the astrometric solution, now typically
negligible thanks to the \textit{Gaia} DR2 catalogue
\citep{gaia2016a,gaia2018,lindegren2018}, to which all the astrometry
was calibrated. For all instruments used in the analysis, an assessment
of the timing accuracy was also included. In most cases, a conservative
timing uncertainty of one second was assumed. When timing biases were
suspected, we only included the cross-track component of the astrometric
position in the astrometric fit, and deweighted the along-track
direction. All acquired and remeasured astrometry is provided in
\rcom{Appendix} Table~\ref{tab:astrometry}.

The peculiarities of \cdthree{}'s outgoing trajectory and, in
particular, its low relative velocity with respect to Earth, kept the
object at small geocentric distances for many weeks after discovery. As
a result, most of the astrometric coverage was obtained when topocentric
parallax was significant, and it is essential to know the precise and
accurate location of the observing telescope, ideally to within a few
meters in the \textit{Gaia} catalog era.  We therefore dedicated
significant effort to obtain accurate coordinates and/or codes for all
the telescopes we used to extract observations of \cdthree{}.

\subsection{Photometric data reduction}

The raw GMOS-N data frames were reduced using standard techniques with
the Gemini DRAGONS Python package \citep[Data Reduction for Astronomy
from Gemini Observatory North and South, ][]{2018ascl.soft11002A}.
Nightly bias frames and twilight flats from the several nights
surrounding the observations were used to create the master bias and
flat-fields. The DAOPHOT software package \citep{stetson1987}, embedded
in the Image Reduction and Analysis Facility \citep[IRAF,
][]{tody1986,tody1993},  was used to perform aperture photometry for all
the GMOS images. The photometry was calibrated to the Sloan Digital Sky
Survey (SDSS) photometric system \cite[$g'$, $r'$ and $i'$,
][]{1996AJ....111.1748F} with the SDSS Data Release catalog 12
\citep{alam2015}, accessed through the SkyServer platform. The resulting
\rcom{individual measurements and errors of} GMOS photometry is provided
in \rcom{Appendix} Table~\ref{tab:photometry}.  \rcom{The resulting
magnitudes in each filter are mean values of individual measurements
with respective filters. That way we diminish the effect of the
brightness variations induced by the rotation of the asteroid.}

The set of images obtained with LDT/LMI for the lightcurve were reduced
using standard bias subtraction and flat field correction from facility
dome flats. \cdthree's photometry was measured using the Photometry
Pipeline \citep{mommert2017}. This pipeline extracted sources with
SourceExtractor using a 3 pixel (1.08") aperture \citep{bertin1996},
astrometrically registered the images based on the \textit{Gaia} DR2
catalog \citep{gaia2018}, and then determined the zero point calibration
for each image by referencing to approximately 50 field stars from the
Pan-STARRS DR1 catalog \citep{flewelling2016}. The photometric
calibration was performed by tying the $VR$-images to the Pan-STARRS1
\rcom{$r_{\rm P1}$ band}.  This technique introduces errors in the
absolute photometric calibration \rcom{as the bands are not identical}
but they are significantly smaller than the typical uncertainty
($\sim0.1-0.2$ magnitude) on the individual measurements. All data
points for the lightcurve are provided in \rcom{Appendix}
Table~\ref{tab:lightcurve}.

\subsection{Calculations of area-to-mass ratio, albedo, density and
phasecurve}

The astrometric data shows that the motion of \cdthree{} is
significantly affected by solar radiation pressure. Establishing its
signature with a $3\sigma$ detection in about three weeks is $2-3$ times
faster than similar analyses in the past. The evolution of the
development of the radiation pressure as a function of time is presented
in Fig.~\ref{fig:amr_evolution}. This improvement is due to the enhanced
precision and accuracy of the astrometry enabled by the \textit{Gaia}
DR2 catalogue \citep{gaia2018,lindegren2018} which permits measuring
individual ground-based positions with 0.05$^{\prime\prime}$ accuracy.

\rcom{In what follows, we interpret the non-gravitational acceleration
as a result of Solar radiation pressure.} Following
\citet{Farnocchia2015_ast4}, we modelled solar radiation pressure
perturbation as a purely radial acceleration $A_1 / r^2$, where $r$ is
the heliocentric distance.  The off-radial components, $A_2$ and $A_3$,
of the Marsden non-gravitational force model
\citep{marsden1969,marsden1973} do not play a significant role in the
albedo-density modelling, unlike for the orbital evolution. The $A_1$
parameter is proportional to the area-to-mass ratio $A/m$ and therefore
can provide useful constraints on the physical properties of the object
and discern between a natural and artificial origin. For a spherical
object, \begin{equation} A_1 = A/m \left(1 + \frac{4}{9}A\right)
\frac{G_S}{c}\ \ ,\ \ A/m = \frac{3}{2 D \rho}, \label{eqn.A1}
\end{equation} where $A$ is the Bond albedo, $G_S$ is the solar
constant, $c$ is the speed of light, $D$ the \rcom{effective} diameter,
and $\rho$ the density \citep{Vokrouhlicky2000,Mommert2014}. \rcom{We
note that this formulation does not take into account the Yarkovsky
effect \citep[cf. ][]{vokrouhlicky1998}, which could contribute to
10-20\% of the total radial non-gravitational acceleration
\citep[e.g.,][]{chesley2014}. Therefore, our calculation is an upper
bound estimate of $A/m$.}

The \rcom{effective} diameter $D$, absolute magnitude $H$, and
\rcom{geometric} albedo $p$ are related by \citep{Pravec2007}: $$ D =
1329\text{ km} \frac{10^{-0.2 H}}{\sqrt{p}}, $$ while the Bond albedo
$A$ is the product of the geometric albedo $p$ and the phase integral
$q$, $$A = p \; q = p \; (0.009082 + 0.4061 G_1 + 0.8092 G_2),$$ where
we have expressed the phase integral $q$ in terms of the $G_1$, and
$G_2$ photometric parameters \citep{muinonen2010}.

\subsection{Orbit computation}

We used a Monte Carlo approach to analyze \cdthree's past trajectory. We
generated 1000 synthetic sets of orbital elements and area-to-mass
ratios by sampling the uncertainty region as calculated from the fit to
the astrometry. We modelled the solar radiation perturbation using all
three coefficients ($A_1, A_2, A_3$) of the Marsden non-gravitational
model \citep{marsden1969,marsden1973}. Given the size of \cdthree{} and
its unknown shape, unlike for the calculation of the area-to-mass ratio,
for orbit computation the off-radial components of the solar radiation
pressure signature are significant on the timescale of the capture
duration. We integrated each synthetic object backwards from 2020 until
the object had been captured into the Earth-Moon system.  The date of
the first perigee within 1 lunar distance (LD) after the insertion into
the Earth-Moon system is used as a proxy for the capture date.

Several synthetic objects's orbits were consistent with a lunar origin
and their distribution at launch from the Moon's surface is provided in
\rcom{Appendix} Fig.~\ref{fig:moonimp} assuming that the Moon is a
sphere of radius 1737 km. In order to trace the possible origin of
\cdthree{} from the Moon, we mapped the outbound trajectories of the
samples originating from the Moon on the Lunar surface. We computed the
state vectors of the samples when leaving the Moon's surface and
transformed them into the Lunar mean Earth/polar axis body-fixed frame
\citep{seidelmann2002} using NASA's Navigation and Ancillary Information
Facility (NAIF) SPICE tools \citep{acton1996,acton2017}.

\section{Results and discussion} \label{s:rd}

\subsection{Physical characterisation}

We used astrometric observations obtained during the apparition to
clearly detect solar radiation pressure acting on \cdthree{} and measure
its area-to-mass ratio, $A/m = (0.65 \pm 0.05) \times 10^{-3}$ m$^2$
kg$^{-1}$. This value implies a natural origin for \cdthree{} because it
is consistent with $A/m$ for other natural objects in the same size
range \citep{micheli2012,micheli2013,micheli2014, Mommert2014,
Mommert2014_MD,  Farnocchia2017} and much lower than typical for
artificial objects \citep{Jenniskens2016}.

\begin{figure} \centering	
	 \includegraphics[width=1.0\textwidth]{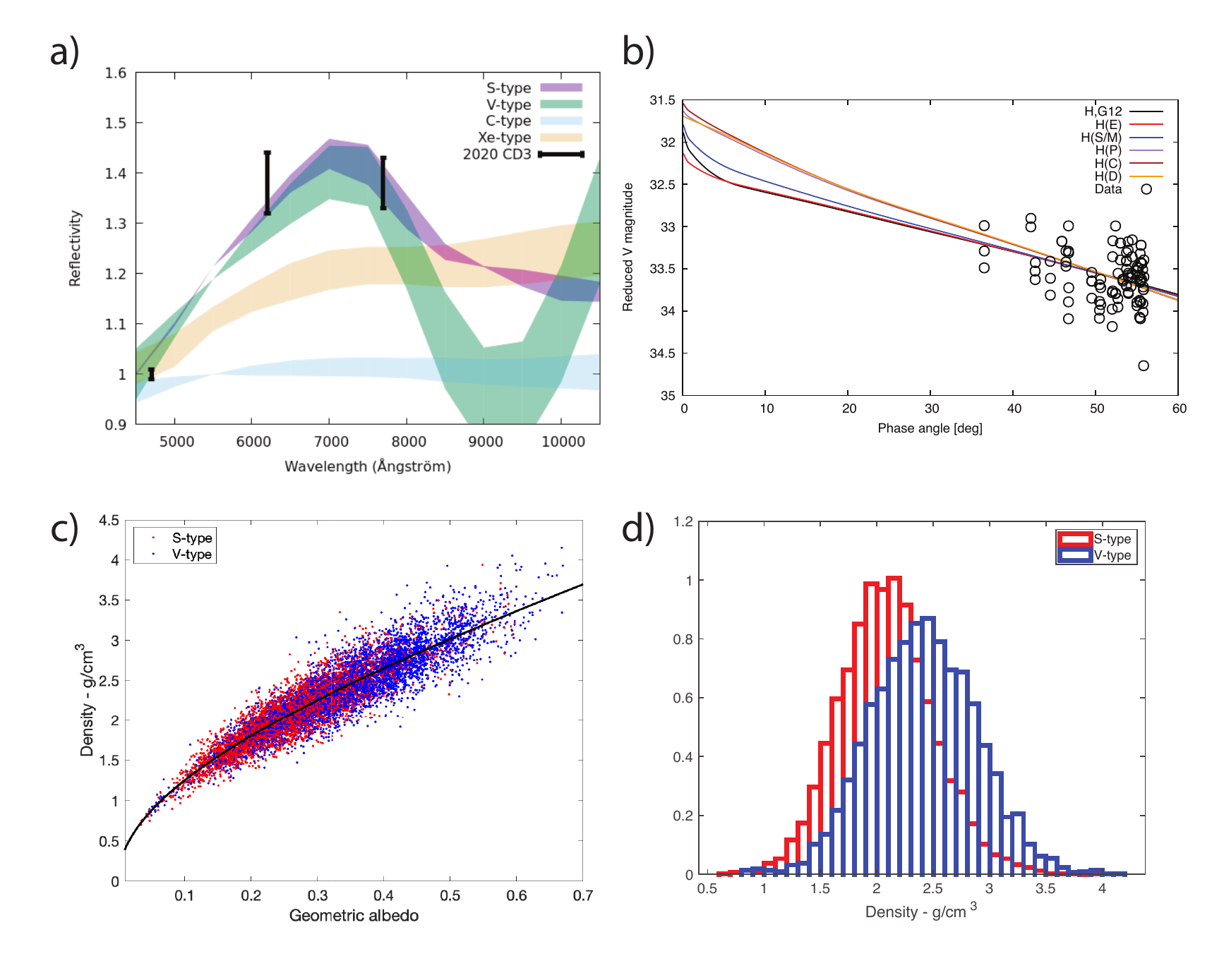} 
	 \caption{\textbf{Physical characterisation of \cdthree{}. (a)}
	 Comparison of the \rcom{color indices} of \cdthree{} obtained with
	 GMOS/Gemini North (black) to the \rcom{reflectance spectra of the}
	 main asteroid taxonomic complexes \rcom{\citep{demeo2009}}.
	 Spectral types C (azure) and Xe (nude, the reddest member of the
	 X-complex) can be ruled out, leaving types S (violet) and V
	 (turquoise)  as plausible choices. \textbf{(b)} \rcom{Photometric
	 phasecurve and constrained photometric fits for \cdthree{}.
	 $H,G_{12}$ is the nonlinear, constrained two-parameter fit, and
	 $H(\ldots)$ refer to one-parameter fits were the slope parameters
	 have been fixed to typical values for different spectral types
	 (E,S/M,P,C,D).} \textbf{(c)} Scatter plot of albedo and bulk
	 density. Red and blue dots correspond to our Monte Carlo
	 distributions for the S and V taxonomic classes, respectively.  The
	 black curve corresponds to the best-fit to the density as a
	 function of the albedo. \textbf{(d)}	 Monte Carlo distribution
	 of the density of \cdthree{} for the S (red) and V (blue) taxonomic
	 classes.} \label{fig:char} \end{figure}

The derived photometric colors ($g'-r' = 0.8 \pm 0.1$, $r'-i' = 0.15 \pm
0.05$) support \cdthree's natural origin as we do not detect extreme
reddening which is associated with artificial objects \citep{miles2011}.
Our broadband photometry suggests that \cdthree{} belongs to the group
of silicate asteroids (Fig.~\ref{fig:char}a), i.e., to the S or V
complexes in the asteroid taxonomy \citep{demeo2013}. Based on physical
characterization alone, we cannot exclude that \cdthree{} is lunar
ejecta, as lunar colors are similar to those of V-type asteroids. The C-
and X-complexes, however, can be ruled out.

We also extracted low-precision \textit{Gaia} $G$-band photometry
\citep{jordi2010} from our astrometric observations to derive the
photometric phase curve and used it in an independent approach to
constrain the spectral classification. The observations of \cdthree{}
are limited to phase angles $36^{\circ}<\alpha<56^{\circ}$ so that the
backscattering region is not covered at all. The poor phase curve
coverage does not allow for the photometric data to be fit with the
standard $(H_V,G_1,G_2)$ system in linear brightness space
\citep{muinonen2010}. Instead, we resort to the alternative technique of
fitting for $(H_V,G_{12})$ in nonlinear magnitude space
\citep{2016P&SS..123..117P} where $G_{12}$ is forced to stay
non-negative and, thus, physically meaningful. The nominal solution,
after converting $G_{12}$ to $(G_1,G_2)$ and $G$ band to $V$ band
assuming $V-G=0.2$, is
$(H_V,G_1,G_2)=(31.88^{+0.03}_{-0.05},0.0^{+0.10}_{-0.0},0.535^{+0.0}_{-
0.069})$ (Fig.~\ref{fig:char}b). We note that the formal uncertainty
estimate for $G_{12}$ is meaningless because its nominal value is a
result of forcing it to be non-negative and the above uncertainty
estimates have been obtained by bootstrapping.

Assuming characteristic slope parameters $(G_1,G_2)$ for different
asteroid taxonomic types  \citep{shevchenko2016} and fitting only for
$H_V$ we find better fits when using slope parameters typical for E, S
and M types than for P, C, and D types (Fig.~\ref{fig:char}b and
Appendix Table \ref{tab:hg1g2fits}). Fixing the slope parameters and
fitting only for $H_V$ results in lower values for the Bayesian
Information Criterion than fitting for both $H_V$ and $G_{12}$
suggesting that the amount of data is not necessarily sufficient for a
meaningful $H_V,G_{12}$ fit let alone a full $H_V,G_1,G_2$ fit. The fit
is also consistent with slope parameters typical for asteroid (4) Vesta
\citep{gehrels1967,shevchenko2016}, the most prominent member of V-type
asteroids. These results are in excellent agreement with the photometric
colors.

In an alternative, synoptic, approach when fitting the radial
component $A_1$ to photometry, the fit to the photometric phase curve
results in an absolute magnitude $H_V=31.9 \pm 0.8$ for \cdthree{}.
The value is consistent with the purely photometric fit, but the
error estimates are more conservative. Assuming the distribution of
possible values of the geometric albedo  \citep[$p_V = 0.26 \pm
0.08$ for S-types, and $0.34 \pm 0.11$ for V-types, ][]{Mainzer2012}
and the phase curve fit for the $H$ magnitude for S- or V-class
asteroids, we obtain a diameter of $1.2^{+0.4}_{-0.2}$ m, one of the ten
smallest NEOs ever found as of 2020 August 10, and amongst the best
characterized with colors, rotation period, and AMR. The derived size is
consistent with the non-detection of \cdthree{} by the Arecibo radar
assuming a non-metallic material composition, excluding an artificial
body or an M-type asteroid (Patrick Taylor, personal communication).

Thus, all our evidence suggests that \cdthree{} is of spectral type S or
V. Although little is known of the color distribution of meter-class
asteroids, our result is consistent with the observed taxonomic
distribution of NEOs with diameters <200 meters where S-class objects
dominate \citep{binzel2019}. Furthermore, it is consistent with
extrapolations of the asteroid taxonomic and orbital element
distribution to small NEOs on Earth-like orbits, the minimoon source
population, which suggest that for $H\sim24.5$, corresponding to S-type
asteroids $\sim40$~m diameter, S-types make up about 40\% of the
population \citep{Jedicke2018-ISRU}.

\begin{figure} \centering
    \includegraphics[width=1.0\textwidth]{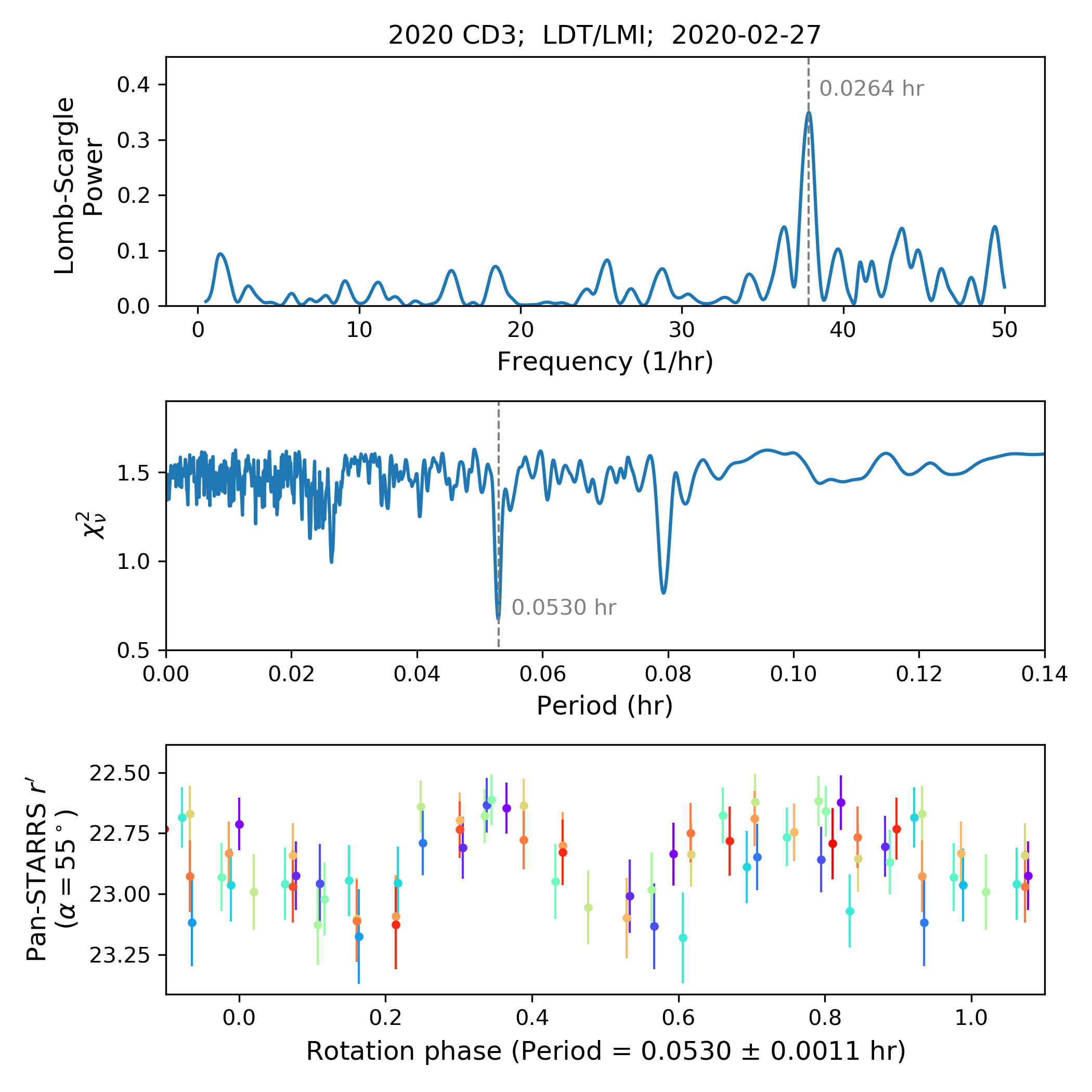}
    \caption{\textbf{Lightcurve of \cdthree{}. Top panel: } the
    Lomb-Scargle periodogram for the rotation period of \cdthree{}. With
    30 s exposures taken over a span of 60 minutes, the range of
    periodogram frequencies is limited from 70 seconds to 2 hours. The
    peak in the periodogram is at 0.0264 h.  \rcom{\textbf{Middle panel:
    } reduced $\chi^2$ residuals from lightcurve curve fits using a 3rd
    order Fourier series. The best fit at 0.0530 hours corresponds to
    the full rotational period.} \textbf{Bottom panel: } LDT lightcurve
    photometry of \cdthree{} calibrated to the Pan-STARRS $r_{\rm P1}$
    filter. The 1-hour sequence of data has been phase folded to the
    best fit period of 0.0530 h (3.2 minutes). The color scale
    represents the ordering of the measurements from the beginning of
    the sequence in blue, to the end in red.  } \label{fig:lc}
    \end{figure}

The lightcurve of \cdthree{}, despite its relatively low signal-to-noise
ratio, shows a strong peak at 0.026 h in a Lomb-Scargle periodogram
and a clear minimum in $\chi^2$ residuals from Fourier fits to the
data at a period = 0.0530 h (Fig.~\ref{fig:lc}). These reduced $\chi^2$
residuals (normalized by the degrees of freedom) were computed for 3rd
order Fourier fits across a range of periods from 0.0001 to 2 h at a
step size of 0.0001 h. Second and fourth order Fourier series produced
overall higher $\chi^2$ values. An approximate 1-sigma error on the
period of 0.0011 h was estimated as the full width at half maximum of
the deepest minimum in the $\chi^2$ plot. Phasing the data to periods at
the limits of this uncertainty range resulted in clear decoherence of
the periodic signal. The best-fit  period $0.0530 \pm 0.0011$ h is
consistent with the Lomb-Scargle periodogram. In particular, given the
apparent $\sim 0.5$ magnitude peak-to-peak amplitude, the second-order
harmonic ($P=0.053$ h) is the most probable interpretation of the
Lomb-Scargle peak for data obtained at a phase angle of $55^{\circ} $
\citep{butkiewicz-bak2017}. We note that the best-fit rotational period
is shorter than the individual integration times of the color
photometry. Therefore, the brightness variation due to the rotation of
\cdthree{} is averaged out in individual photometric color
measurements.  Assuming a double-peaked lightcurve, a period of about
3.2 minutes ($\sim190$~s) is a reasonable interpretation, however due to
the low signal-to-noise of these data, the period is not strongly
constrained. Non-principal axis rotation cannot be ruled out with the
available data. The observed rotational period is at least an order of
magnitude slower than the predicted mean value from a Maxwellian
rotational distribution for meter-sized objects \citep{bolin2014}. This
implies that radar may be better suited for the detection of minimoons
than had been previously anticipated, because the radar signal is
smeared less by asteroid rotation than suggested by extrapolations of
size-rotation-rate models.

The two unknowns in Eq. \ref{eqn.A1} are the albedo and density but they
are constrained by the other measured parameters.  Given that our
photometric model implies that \cdthree\ is either an S or V type
asteroid, we generated synthetic albedos for $10\,000$ of each type
according the type-specific albedo distributions of \citet{Mainzer2012}.
 Similarly, we generated the same number of random $A_1$ values using a
normal distribution with a mean and width given by the central value and
uncertainty on our measured $A_1 = (3.1 \pm 0.2) \times 10^{-9}$ m
s$^{-2}$.  The pairs of synthetic albedo-$A_1$ values were then used to
calculate the object's density (Fig.~\ref{fig:char}c-d).  For the S-type
assumption we find $\rho = 2.1\pm 0.4$~g cm$^{-3}$ whereas for the
V-type assumption $\rho = 2.4\pm 0.5$~g cm$^{-3}$.  In both cases, the
inferred density is consistent with typical asteroid densities
\citep{Carry2012}. We note that the possible effect of the
Yarkovsky force can potentially increase the estimated density values by
10-20\% \citep{chesley2014} so that our density estimates represent the
lower bound of values. However, this does not have a major impact on the
interpretation of the results.

Thus, our physical characterization of \cdthree{} indicates that it is
a silicate  body, perhaps a free-floating analogue of what appear to be
monolithic boulders found on the surface of larger asteroids such as
(25143) Itokawa, the S-type asteroid investigated in-situ by the
\textit{Hayabusa} spacecraft \citep{Saito2006}. Alternatively, it could
be a small rubble-pile aggregate more like the 2008~TC$_3$
\citep[e.g.][]{Jenniskens2009-TC3}. While the internal structure of
meter-scale asteroids is currently unknown we expect that favourable
appearances of small NEOs and future minimoons will provide more
opportunities for detailed characterization of these small asteroids.

\subsection{Orbital evolution} \label{sss.OrbitalEvolution}

\begin{figure} \centering	
    \includegraphics[width=1.0\textwidth]{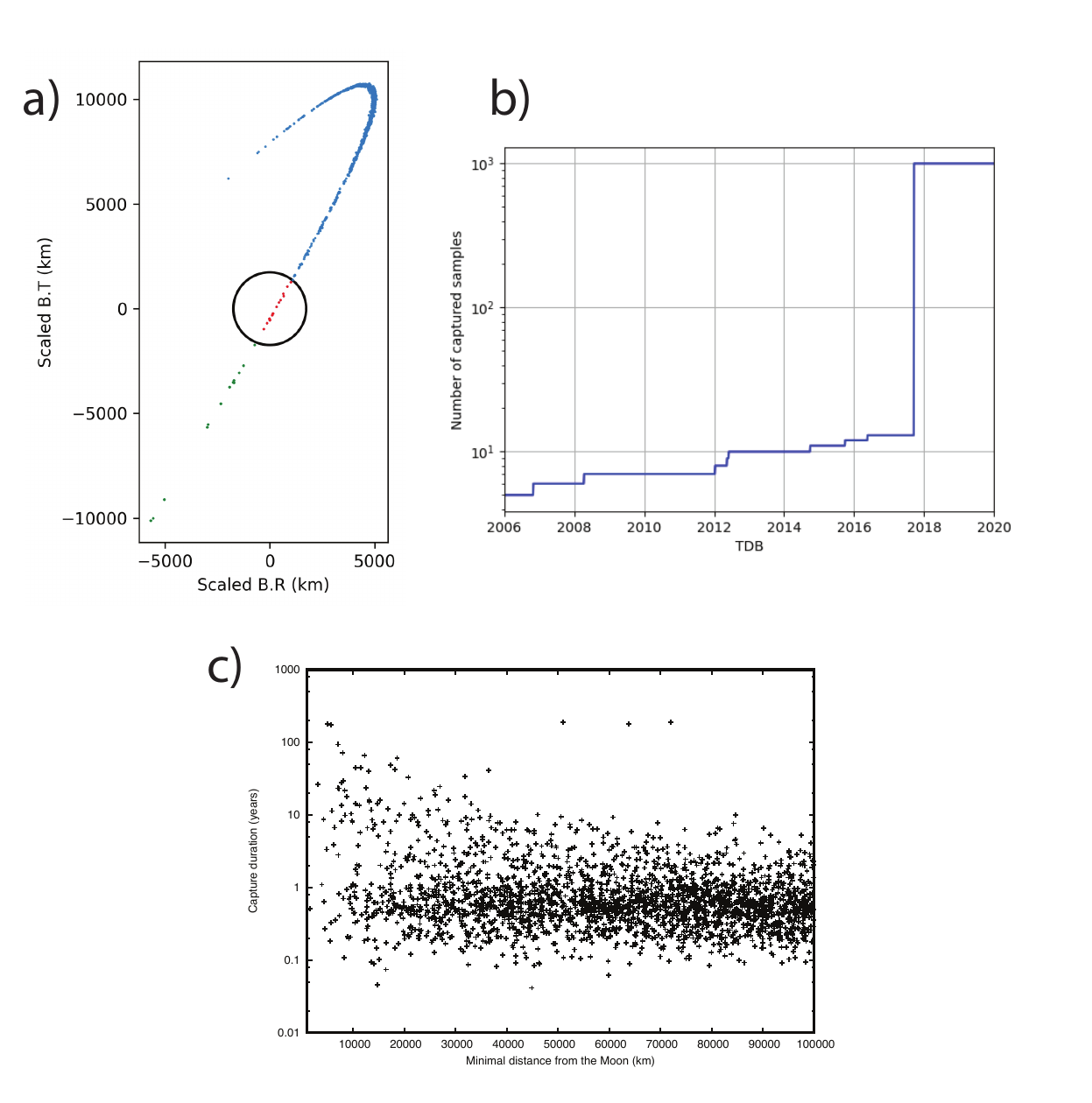}
    \caption{\textbf{Orbital evolution of \cdthree{}. (a)} Monte Carlo
    samples of the \cdthree{} trajectory mapped onto the outbound scaled
    B-plane \citep{farnocchia2019} of the Moon (black circle) on 2017
    September 15. On that date there was a close encounter between
    \cdthree{} and the Moon which results in a trivergence of orbital
    solutions when going further back in time:  (blue) samples captured
    on a geocentric orbit during this encounter, (red) samples
    originating from the Moon, and (green) samples remaining in an orbit
    around  Earth prior to the encounter. \textbf{(b)} The cumulative
    number of Monte Carlo \rcom{samples} bound to the Earth-Moon system
    as a function of time when integrating backwards from 2020. 
    \textbf{(c)} The capture duration of minimoons as a function of
    their minimum lunacentric distance \citep[using data
    from][]{fedorets2017}. A close encounter between a minimoon and the
    Moon typically increases the capture duration.}
    \label{fig:moonbplane} \end{figure}

Integrating \cdthree's trajectory into the past indicates that it was
bound to the Earth and it's orbit was deterministic after a close
approach to the Moon on 2017 September 15 (Figure
\ref{fig:moonbplane}a).  Prior to this encounter there are three
possible behaviors: 1) $97.3\%$ escape the Earth-Moon system,
corresponding to a scenario in which this encounter is responsible for
the capture of \cdthree{} by the Earth-Moon system; 2) $1.4\%$ intersect
the Moon's surface, which corresponds to the hypothesis that \cdthree{}
is lunar ejecta; 3) $1.3\%$ remain in Earth orbit (potentially for more
than 10 years). Therefore, we conclude that \cdthree{} was in orbit
around Earth since at least 2017 September 15. Since then, it completed
11 orbits around the Earth with intervals between successive perigees of
70 to 90 days. Its minimum geocentric distance was between $12\,900$ and
$13\,400$ km on 2019 April 4 and it escaped the Earth's Hill sphere
($\sim$0.01 au) on 2020 March 7 after a final perigee on 2020 February
13 at a geocentric distance of about $47\,000$ km.  Oddly, it passed its
last perigee just two days before its discovery. \cdthree{}'s Earth-like
orbit means it has a long synodic orbital period so it will not
approach Earth again until March 2044 at about 10 lunar distances, well
outside Earth's Hill sphere.

The capture duration of \cdthree{} of at least 2.7 years
(Fig.~\ref{fig:moonbplane}b) may seem exceptionally long considering
that orbital simulations suggest that the average capture duration of
minimoons is about nine months \citep{fedorets2017}. However,
there is an inverse correlation between the average capture
duration and the minimum lunacentric distance when the encounter
distance is less than $30\,000$~km (Fig.~\ref{fig:moonbplane}c). In such
cases, minimoons may become captured for years or even decades. Although
only $\sim2$\% of minimoons have capture durations greater than three
years those objects' total capture duration time is 23\% of the
cumulative capture duration time of all simulated minimoons. Based on
the close encounter of \cdthree{} with the Moon, it is not surprising
that \cdthree{} undergoes a longer geocentric capture than an average
minimoon. The distribution of possible capture durations of \cdthree{}
is thus in agreement with theoretical predictions
(Fig.~\ref{fig:moonbplane}b).

We argue that a lunar origin for \cdthree{} is highly unlikely (see
Subsect. \ref{su:lunar}) and therefore assume that the object originated
in the main asteroid belt.  Based on its pre-capture heliocentric orbit
it has a $(72\pm1)$\% probability of having been ejected by the $\nu_6$
secular resonance with, primarily, Saturn
\citep{Granvik2018-Icarus-NEO-model}.  A provenance in the inner main
belt would also favor its identification in the S-type taxonomy since
S-types dominate that region of the belt.  There is a $(28\pm1)$\% for
it having originated in the Hungaria region and a negligible
$(0.5\pm0.03)$\% probability that it was ejected from the outer region
of the main belt by the 3:1 mean-motion resonance with Jupiter.  The
reported uncertainties on the probabilities are the standard error on
the mean across several discrete cells in the
\citet{Granvik2018-Icarus-NEO-model} NEO population model. An inner belt
source for \cdthree{} is in agreement with a silicate-rich asteroid
composition which is dominant in that region
\citep{DeMeo2014-Nature-MainBeltMapping}.

\subsection{Exploring the lunar ejecta hypothesis} \label{su:lunar}

There is a possibility that \cdthree{} could have been spall ejected by
a recent lunar impact event (Subsect.~\ref{sss.OrbitalEvolution}) and we
assess the likelihood of this scenario by examining the contemporary
production rate of small craters on the Moon.

The largest crater to form annually on the Moon is approximately 50 m in
diameter based on a survey of fresh impact craters identified on the
Moon using "before" and "after" images from the Lunar Reconnaissance
Orbiter (LRO) Narrow Angle Camera (NAC) \citep{speyerer2016}.   
Accordingly, if \cdthree\ was launched from the lunar surface on 2017
September 15 a crater of this scale would need to be capable of
launching a meter-sized minimoon off the Moon and onto the trajectory
described above.

An asteroid striking the Moon creates a crater approximately 20 times
its own size \citep{melosh1989} so a 2.5~m diameter projectile can make
a 50 m diameter crater. \citet{hirase2004} investigated the relationship
of ejecta velocity relative to the ejecta-to-impactor diameter ratio in
laboratory experiments, an analysis of secondary craters produced by
lunar and Martian craters, and ejecta from the asteroid (4) Vesta that
make up the Vesta family (often called Vestoids).  At an ejecta/impactor
diameter ratio of $\sim0.4$, corresponding to the ejection of a 1~m
diameter minimoon by a 2.5~m diameter projectile, the typical ejection
speed are a few tens of m~s$^{-1}$ and certainly $<100$~m~s$^{-1}$ ---
much smaller than lunar escape velocity ($\sim 2\,400$~m~s$^{-1}$). 
Indeed, the results of \citet{hirase2004}'s analysis suggest that
launching a 1~m diameter minimoon off the lunar surface requires the
impact of a km-scale asteroid, an unlikely event that surely would have
been noticed on or soon after 2017 September 15.  Furthermore, the
population of NEOs is $\gtrsim90$\% known at this time and no impacts
were predicted on that date.  Accordingly, we reject a lunar ejecta
origin for \cdthree. In summary, while NEO-based models
\citep{granvik2012,fedorets2017} indicate that an annual capture of a
meter-sized asteroid is likely, the production of similar-sized lunar
ejecta at the same rate can be ruled out. Hence, minimoon capture from
the NEO population is a dominating mechanism for maintaining the
minimoon steady-state population.

An additional blow to the lunar origin hypothesis for \cdthree\ comes
from lunar meteorites that were blasted off the Moon in the past. 
\citet{warren1994}, whose analysis builds on the work of
\citet{melosh1985}, argues that most lunar meteorites came from lunar
craters that were hundreds of meters to several km in diameter and that
the meteoroid precursor bodies to the meteorites were 2-10 cm in
diameter prior to entering Earth's atmosphere. Lunar meteorite cosmic
ray exposure ages indicate that only about half took less than
$100\,000$ years to get to Earth \citep{warren1994}.  Given that
minimoon orbital lifetimes are typically on the order of a year it
implies that those meteoroids spent most of their time on heliocentric
orbits before being delivered back to Earth, not in the Earth-Moon
system.  Taken together, it suggests that it is difficult for small
craters to launch sizeable bodies off the Moon; if small impact events
could do so, we might expect very young lunar meteorites to dominate the
fall and find record on Earth.

We emphasize that the impact capable of producing an ejecta of the size
of \cdthree{} would have been very bright. Moreover, the distribution of
the subset of sample orbits originating from the Moon point the majority
of them to the part of the dark size of the Moon facing towards Earth,
providing optimal observing conditions (Appendix
Fig.~\ref{fig:moonimp}). No major impacts have been reported, including
the NELIOTA telescope \citep{xilouris2018}, the NASA lunar impact
monitoring (William Cooke, private communication). Moreover, no reports
of a new km-sized craters on the Moon have been announced.

In summary, we consider the lunar origin of \cdthree{} to be extremely
unlikely.

\subsection{Detectability of \cdthree{}}

\begin{figure} \centering \includegraphics[trim=10 275 10
10,clip,width=0.95\textwidth]{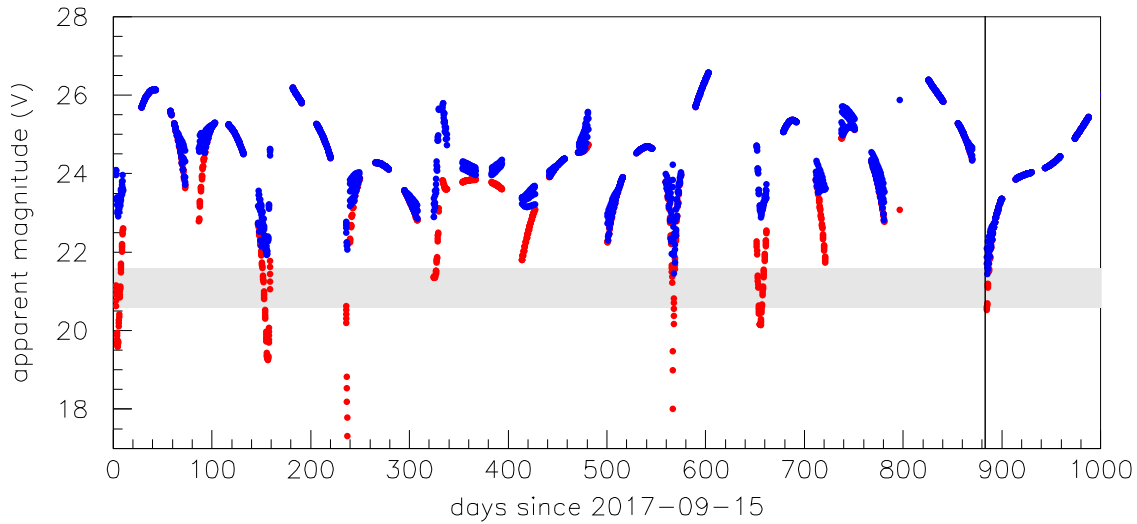}
\caption{\textbf{Detectability windows for \cdthree{}.}  The red points
represent \cdthree's V-band apparent magnitude every 2 hours over a
period of about 2.7 years beginning on the date of its close approach to
the Moon on 2017 September 15. The blue points represent \cdthree's
V-band apparent magnitude after accounting for trailing losses specific
to the discovery observatory. To mimic CSS's observing strategy, the
points are only shown when \cdthree\ is more than $60^{\circ}$ from the
Sun, more than $45^{\circ}$ from the Moon when it is $<50$\%
illuminated, and from 4~h to 12~h UTC. The vertical line is on the date
on which \cdthree was discovered (2020 February 15). The horizontal gray
band represents $\pm0.5$~magnitudes centered on the average limiting
magnitude of the CSS Mt.~Lemmon telescope.  Thus, the only time when
that telescope could detect \cdthree\ is when blue dots appear in or
below the gray band.   
} \label{fig:G96-Vtrail-vs-day} \end{figure}

The discovery of \cdthree{} occurred at the last window of opportunity
(Fig.~\ref{fig:G96-Vtrail-vs-day}). However, simulations by
\citet{fedorets2020}, and the fact that \RH was discovered only three
months into its captured time period of one year, suggest that the
last-minute discovery of \cdthree\ is not a typical situation. During
the undisputed capture period of 2.7 years there were six distinct
intervals during which \cdthree{} was brighter than the discovery
observatory's (CSS's Mt.~Lemmon) limiting magnitude
(Fig.~\ref{fig:G96-Vtrail-vs-day}). It even briefly reached $V<16$ when
it approached to within about $20\,000$~km, below the orbits of
geosynchronous satellites.  The problem is that during the detectability
windows, when it was bright and close to Earth, it also had a high
apparent rate of motion so that it would have left a trailed image on
the detector, spreading out the light from the object and reducing the
per pixel signal-to-noise ratio (SNR) to a level below the system's
detection threshold. Taking these trailing losses into account, there
were only three 2-hour time segments during the entire 2.7~years in
which \cdthree\ was detectable by the Mt.~Lemmon telescope,
corresponding to $\sim0.03$\% of the time under the best of
circumstances.  A similar analysis for the Pan-STARRS1 telescope
\citep{chambers2016panstarrs1} finds that there were only four short
time periods during which it could have detected \cdthree. Pan-STARRS1
reaches a fainter limiting magnitude than Mt.~Lemmon due to its larger
aperture and better seeing statistics but its smaller pixels makes it
less sensitive to fast-moving objects like minimoons.

\subsection{Minimoon population}

We expect there to exist a much larger but undiscovered population of
minimoons that are similar or smaller in size to \cdthree
\citep{fedorets2017,granvik2012} --- they are just difficult to detect
due to their faintness, rate of motion, and infrequent windows of
observational opportunity (Fig. \ref{fig:G96-Vtrail-vs-day}). 
Estimating the minimoon populations size-frequency distribution by
debiasing the discovered population of two objects is essentially
impossible given that they are so difficult to detect and were at the
limit of the system's detection capability. In addition to the two 
minimoons discovered by CSS, observations of meteors created by
meteoroids that had a high probability of being geocentric prior to
entering the atmosphere \citep{clark2016,shober2019} support the
existence of a minimoon population. These meteor observations are also
difficult to convert into a minimoon population estimate because a
meteor's apparent brightness, in both the optical and radar, is
dominated by the meteoroids diameter and its speed.  Since meteors
generated by minimoons have the lowest possible meteor speed,
essentially equal to Earth's escape speed, they are the faintest
possible meteors.  Thus, in order for them to be bright enough to be
detected they must be large and therefore rare.

To quantify the detection frequency of minimoons we apply
\citet{bolin2014}'s modelling of the performance  of Pan-STARRS1 survey
\citep[PS1;][]{chambers2016panstarrs1} to the CSS's Mt. Lemmon
observatory that discovered both of the telescopically identified
minimoons.  The application is appropriate because the two observatories
have roughly similar capabilities, especially considering all the
difficulties involved in modeling the detection of faint, fast-moving
minimoons, and the statistics of just two objects. The modelled PS1
survey has a peak probability of detecting minimoons at $H_V=31.5 \pm
1.5$ so Mt. Lemmon's discovery of \cdthree\ with $H_V=31.9 \pm 0.8$ is
not surprising.  Furthermore, \citet{bolin2014} estimated that PS1 (and
therefore Mt. Lemmon) could detect about $0.01$ minimoons per lunation
or about one every $\sim$8.1~years as compared
to the $\sim$14~year interval since CSS's discovery of \RH.  We think
the $\sim2\times$ discrepancy in the time interval is not significant
given that 1) \citet{bolin2014} used the earlier and larger minimoon
size-frequency distribution of \citet{granvik2012} compared to the
revised distribution of \citet{fedorets2017} and because 2) it is
intrinsically difficult to model discovery rates at the limits of
detectability in both flux and rate of motion (see
Fig.~\ref{fig:G96-Vtrail-vs-day}). Moreover, assuming Poisson-like
discovery statistics, and that the CSS Mt. Lemmon survey has been in
operation at roughly the same capability level for 20~years, \rcom{over
the same period} there is a $\sim68$\% probability of discovering $\le2$
minimoons. 
Therefore, the discovery of \cdthree{} 14 years after the discovery of
\RH, a minimoon with $H_V=29.9\pm0.3$, is in line with the capture
frequency of minimoons predicted by existing population models, and
consistent with their predicted discovery rate \citep{bolin2014}.

An additional complication in debiasing the minimoon population
identified in asteroid surveys is the difficulty of identifying rare
natural objects among numerous artificial ones \citep{jedicke2018}.  As
sky surveys have become more powerful and efficient at identifying faint
and trailed objects they have been detecting ever more artificial
geocentric objects, often on minimoon-like orbits. Distinguishing both
\cdthree\ and \RH\ from artificial objects upon their discovery was
initially inadvertently affected by human biases that objects on
geocentric orbits are artificial and correcting the observation
statistics for this bias will be difficult.

\section{Conclusions} \label{s:conclusions}

We provide an in-depth study of the orbital and physical characteristics
of Earth's second discovered minimoon, \cdthree{}. The combination of
its area-to-mass ratio derived from its solar radiation pressure
signature, its reflectance spectrum that is consistent with silicate
asteroids, and its non-detection by radar provides evidence that
\cdthree{} is a natural object. Its derived geometric albedo - bulk
density distributions are consistent with values typical of silicate
asteroids. \cdthree{} could be a free-floating silicate analogue of
boulders discovered on surfaces of larger asteroids.

High precision astrometry enabled by the \textit{Gaia} mission permits
the measurement of ground-based astrometry of asteroids to a level of
0.05$^{\prime\prime}$ in the best cases. This, in turn, provides better,
faster constraints on the solar radiation pressure signature for small
asteroids. For the best possible results, special attention needs to be
paid to 1) using accurate and precise geographical coordinates for
ground-based observatories and 2) time-keeping.

The geocentric orbital evolution of \cdthree{} includes a close
dynamical interaction with the Moon on 2017 September 15. It was bound
on a geocentric orbit for at least 2.7 years, which makes it an
exceptionally long capture compared to typical capture duration results
from simulations. However, a long duration is actually typical for
minimoons that experience close approaches to the Moon.  Some orbital
trajectories suggest a lunar ejecta  origin of \cdthree{}, but we showed
that this possibility is unlikely.

The discovery of \cdthree{} is in line with the most recent theoretical
predictions for the steady-state population of minimoons
\citep{fedorets2017}, supporting the prediction for an increased
discovery rate of minimoons \citep{fedorets2020} with the upcoming Vera
C. Rubin  Observatory's Legacy Survey of Space and Time \citep[LSST;
][]{ivezic2019}. More discoveries of minimoons are also anticipated
through improvements to the ongoing NEO surveys such as CSS
\citep{bolin2014} but trailing losses remain a major factor affecting
the discovery of minimoons. A rapid determination of the nature of
temporarily-captured objects after their discovery remains a challenge
which is expected to become even more pressing with the anticipated
increase in their discovery rate with LSST.

The discovery and characterization of \cdthree{} paves the way for the
observational study of minimoons as a population instead of a curiosity
with serendipitous discoveries, as well as for identifying candidate
targets for cost-effective space missions to these unexplored objects at
the asteroid-meteoroid boundary \citep{granvik2013,jedicke2018}.

\acknowledgements

The authors thank Bryce Bolin, William Cooke, Alan Fitzsimmons,
Tom\'a\v{s} Kohout, Antti Penttil\"a, Patrick Taylor and Anne Virkki for
helpful discussions, \rcom{the anonymous referee for his/her insightful
comments}, and Daniel Hestroffer for his hospitality during the research
visit of G.F. to the Paris Observatory.

Based on observations made with the Nordic Optical Telescope, operated
by the Nordic Optical Telescope Scientific Association at the
Observatorio del Roque de los Muchachos, La Palma, Spain, of the
Instituto de Astrofisica de Canarias. The data presented here were
obtained in part with ALFOSC, which is provided by the Instituto de
Astrofisica de Andalucia (IAA) under a joint agreement with the
University of Copenhagen and NOTSA.

Based on observations obtained with MegaPrime/MegaCam, a joint project
of CFHT and CEA/DAPNIA, at the Canada-France-Hawaii Telescope (CFHT)
which is operated by the National Research Council (NRC) of Canada, the
Institut National des Sciences de l'Univers of the Centre National de la
Recherche Scientifique (CNRS) of France, and the University of Hawaii.
The observations at the CFHT were performed with care and respect from
the summit of Maunakea which is a significant cultural and historic
site.

The authors acknowledge the sacred nature of Maunakea, and appreciate
the opportunity to obtain observations from the mountain. This work is
partly based on observations obtained at the international Gemini
Observatory, a program of NSF's OIR Lab, which is managed by the
Association of Universities for Research in Astronomy (AURA) under a
cooperative agreement with the National Science Foundation. on behalf of
the Gemini Observatory partnership: the National Science Foundation
(United States), National Research Council (Canada), Agencia Nacional de
Investigaci\'{o}n y Desarrollo (Chile), Ministerio de Ciencia,
Tecnolog\'{i}a e Innovaci\'{o}n (Argentina), Minist\'{e}rio da
Ci\^{e}ncia, Tecnologia, Inova\c{c}\~{o}es e Comunica\c{c}\~{o}es
(Brazil), and Korea Astronomy and Space Science Institute (Republic of
Korea). The observations were obtained as part of Gemini Director's
Discretionary Program GN-2020A-DD-107. The GMOS-N observations were
acquired through the Gemini Observatory Archive at NSF's NOIRLab and
processed using DRAGONS (Data Reduction for Astronomy from Gemini
Observatory North and South).

This work was enabled by observations obtained with the University of
Hawai`i's 2.2-meter telescope and the Gemini North telescope, and the
Canada-France-Hawaii telescope, all located within the Maunakea Science
Reserve and adjacent to the summit of Maunakea. We are grateful for the
privilege of observing the Universe from a place that is unique in both
its astronomical quality and its cultural significance.

These results made use of the 4.3 m Lowell Discovery Telescope at Lowell
Observatory. Lowell is a private, non-profit institution dedicated to
astrophysical research and public appreciation of astronomy and operates
the LDT in partnership with Boston University, the University of
Maryland, the University of Toledo, Northern Arizona University and Yale
University. The Large Monolithic Imager was built by Lowell Observatory
using funds provided by the National Science Foundation (AST-1005313).
Part of the LDT observations have been obtained thanks to the NASA
SSO-NEOO grant number 80NSSC19K1586. Some of the LDT observations were
obtained by the University of Maryland observing team, consisting of
L.~M. Feaga, Q.-Z. Ye, J.~M. Bauer, T.~L. Farnham, C.~E. Holt, M.~S.~P.
Kelley, J.~M. Sunshine, and M.~M. Knight.

This work has made use of data from the European Space Agency (ESA)
mission {\it Gaia} (\url{https://www.cosmos.esa.int/gaia}), processed by
the {\it Gaia} Data Processing and Analysis Consortium (DPAC,
\url{https://www.cosmos.esa.int/web/gaia/dpac/consortium}). Funding for
the DPAC has been provided by national institutions, in particular the
institutions participating in the {\it Gaia} Multilateral Agreement.

Based on the observations obtained with the Chinese Near-Earth Object
Survey Telescope (CNEOST). We thank Bin Li for providing the exposure
catalog.

Based on observations obtained with the Samuel Oschin 48-inch Telescope
at the Palomar Observatory as part of the Zwicky Transient Facility
project. ZTF is supported by the National Science Foundation under Grant
No. AST-1440341 and a collaboration including Caltech, IPAC, the
Weizmann Institute for Science, the Oskar Klein Center at Stockholm
University, the University of Maryland, the University of Washington,
Deutsches Elektronen-Synchrotron and Humboldt University, Los Alamos
National Laboratories, the TANGO Consortium of Taiwan, the University of
Wisconsin at Milwaukee, and Lawrence Berkeley National Laboratories.
Operations are conducted by COO, IPAC, and UW.

The Catalina Sky Survey is funded since 1998 by the National Aeronautics
and Space Administration's Near Earth Object Observations program,
currently under Grant No. 80NSSC18K1130.

The Pan-STARRS1 Surveys (PS1) have been made possible through
contributions by the Institute for Astronomy, the University of Hawaii,
the Pan-STARRS Project Office, the Max-Planck Society and its
participating institutes, the Max Planck Institute for Astronomy,
Heidelberg and the Max Planck Institute for Extraterrestrial Physics,
Garching, The Johns Hopkins University, Durham University, the
University of Edinburgh, the Queen's University Belfast, the
Harvard-Smithsonian Center for Astrophysics, the Las Cumbres Observatory
Global Telescope Network Incorporated, the National Central University
of Taiwan, the Space Telescope Science Institute, and the National
Aeronautics and Space Administration under Grant No. NNX08AR22G issued
through the Planetary Science Division of the NASA Science Mission
Directorate, the National Science Foundation Grant No. AST-1238877, the
University of Maryland, E\"{o}tv\"{o}s Lor\'{a}nd University (ELTE), and
the Los Alamos National Laboratory.

M.M. would like to thank Detlef Koschny, Luca Conversi and Erwin Schwab
for their support with the CAHA Schmidt observations.

G.F. was supported by STFC Grant ST/P000304/1. M.G. was partly supported
by the Academy of Finland. Support for M.D. and N.M. was provided by
NASA NEOO grant NNX17AH06G in support of the Mission Accessible
Near-Earth Object Survey (MANOS). W.B. and R.J. were supported in part
by NASA's Near Earth Object Observations program (Grant Number
80NSSC17K0153). L.B. acknowledges funding from the Science Technology
Funding Council (STFC) Grant Code ST/T506369/1. T.S. is supported by
Gemini Observatory through a Gemini Science Fellowship. Part of this
research was conducted at the Jet Propulsion Laboratory, California
Institute of Technology, under a contract with NASA.

This work made use of NASA's Astrophysics Data System Bibliographic
Services and the data and services provided by the International
Astronomical Union's Minor Planet Center.
(\url{https://minorplanetcenter.net/data}).

\section*{Author contributions}

G.F., M.M., R.J., S.N, D.Fa., M.G., N.M., M.S., R.We., K.W., E.C., Q.Y.,
W.B. wrote the paper. K.W. and T.P. discovered the object. G.F and M.M.
acquired the NOT data, supported by A.D. R.We. and R.Wa. acquired the
CFHT data. N.M., Q.Y., M.D. and M.M. acquired the LDT data. D.F\"{o}.
acquired the UH88 data. G.F., M.S., L.B., M.M., M.G. wrote the Gemini
North DD proposal. M.S., G.F., T.S., D.M.F., J.R., A.S. acquired the
Gemini North data. M.M., R. We., Q.Y. searched for precovery images.
M.M. and D.Fa. computed the area-to-mass ratio. N.M. derived the
rotational period. D.Fa., M.M., M.G., R.J., G.F. physically
characterised the object. S.N., D.Fa. and M.M. calculated the orbit.
W.B. and R.J. investigated the lunar ejecta hypothesis. M.G. estimated
the source region of the object. R.J., E.C., R.We. and M.M. investigated
the detectability of the object. R.J., G.F., M.M., M.G., D.Fa.
interpreted the results.

\software{DAOPHOT (Stetson 1987), IRAF (Tody 1986, 1993), NASA NAIF
SPICE tools (Acton 1996, Acton et al. 2018), Sextractor (Bertin \&
Arnouts 1996), DRAGONS (AURA Gemini Observatory-Science User Support
Department 2018), Photometry Pipeline (Mommert 2017).}

%%%%%%%%%%%%%%%%%%%% REFERENCES %%%%%%%%%%%%%%%%%%

\bibliographystyle{aasjournal}

\bibliography{bibliography}

%%%%%%%%%%%%%%%%% APPENDICES %%%%%%%%%%%%%%%%%%%%%

\appendix

\begin{longtable}[h]{|l|l|l|l|l|} \caption{Telescopes used in this work
and their purpose.} \\
%    \begin{tabular}{|l|l|l|l|l|l|}
Telescope        & Aperture (m)  &  Astrometry     &  Photometry & 
Lightcurve  \\ \hline \endfirsthead \multicolumn{4}{r}{\textit{Appendix
Table \ref{tab:telescope} continued}} \\ \hline \endhead
\multicolumn{4}{r}{\textit{Appendix Table \ref{tab:telescope} continued
on next page}} \\ \endfoot \endlastfoot \hline CSS Mt. Lemmon           
      & 1.5           & \checkmark           &             &            
 \\ Calar Alto Schmidt              & 0.8           & \checkmark        
  &             &              \\ Nordic Optical Telescope        & 2.5 
         & \checkmark                              &             &      
       \\ Gemini North                    & 8.1           &             
                           &    \checkmark   &              \\
Canada-France-Hawai'i Telescope & 3.6           &  \checkmark           
                 &             &              \\ Lowell Discovery
Telescope      & 4.3           &   \checkmark                           
 &             &  \checkmark   \\ U. of Hawai'i 2.2-meter         & 2.2 
        & \checkmark                             &             &
\label{tab:telescope} \end{longtable}

\begin{longtable}[h]{|l|l|l|l|l|l|l|l|} \caption{Newly acquired or
remeasured astrometric data for \cdthree{}. The columns are, in order
from left to right: observation date (UTC); right ascension;
declination; \textit{Gaia} system magnitude \citep{jordi2010}; MPC
observatory code; right ascension uncertainty in arcseconds; declination
uncertainty in arcseconds; telescope name.} \\
%    \begin{tabular}{|l|l|l|l|l|l|l|l|}
 Date (UTC)        & $\alpha$     &  $\delta$    & G    & Code & 
 $\sigma_{\alpha}$ & $\sigma_{\delta}$ & Telescope \\ \hline
 \endfirsthead \multicolumn{4}{r}{\textit{Appendix Table
 \ref{tab:astrometry} continued}} \\ Date (UTC)        & $\alpha$     & 
 $\delta$    & G    & Code &  $\sigma_{\alpha}$ & $\sigma_{\delta}$ &
 Telescope \\ \hline \endhead \multicolumn{4}{r}{\textit{Appendix Table
 \ref{tab:astrometry} continued on next page}} \\ \endfoot \endlastfoot
 2020 02 15.526427 & 13 03 37.570 & +09 17 38.40 & 19.6 & G96 & 0.11  &
 0.16 & CSS Mt. Lemmon \\ 2020 02 17.975500 & 13 45 24.910 & +19 18
 44.71 & 21.0 & Z84 & 0.30  & 0.30 & Calar Alto Schmidt \\ 2020 02
 17.984601 & 13 45 26.960 & +19 20 34.07 & 20.9 & Z84 & 0.32  & 0.32 &
 Calar Alto Schmidt \\ 2020 02 17.993701 & 13 45 28.601 & +19 22 22.81 &
 21.1 & Z84 & 0.21  & 0.21 & Calar Alto Schmidt  \\ 2020 02 18.002805 &
 13 45 29.872 & +19 24 10.44 & 21.0 & Z84 & 0.22  & 0.22 & Calar Alto
 Schmidt  \\ 2020 02 18.011907 & 13 45 30.718 & +19 25 57.10 & 21.3 &
 Z84 & 0.17  & 0.17 & Calar Alto Schmidt  \\ 2020 02 18.021008 & 13 45
 31.196 & +19 27 42.04 & 21.3 & Z84 & 0.19  & 0.19 & Calar Alto Schmidt 
 \\ 2020 02 21.093805 & 14 03 50.575 & +24 20 51.66 & 21.8 & Z84 & 0.24 
 & 0.24 & Calar Alto Schmidt  \\ 2020 02 21.114761 & 14 03 44.360 & +24
 22 47.25 & 21.8 & Z84 & 0.27  & 0.27 & Calar Alto Schmidt  \\ 2020 02
 21.177717 & 14 03 22.794 & +24 27 37.51 & 21.8 & Z84 & 0.23  & 0.23 &
 Calar Alto Schmidt  \\ 2020 02 21.198677 & 14 03 15.400 & +24 28 53.86
 & 21.7 & Z84 & 0.15  & 0.15 & Calar Alto Schmidt  \\ 2020 02 21.174711
 & 14 03 52.507 & +24 31 07.79 & 22.0 & Z23 & 0.13  & 0.15 & NOT        
      \\ 2020 02 21.199832 & 14 03 42.272 & +24 33 04.85 & 21.6 & Z23 &
 0.10  & 0.12 & NOT             \\ 2020 02 21.203876 & 14 03 40.597 &
 +24 33 22.12 & 21.6 & Z23 & 0.07  & 0.07 & NOT              \\ 2020 02
 24.586055 & 14 14 38.369 & +27 28 39.86 & 22.6 & 568 & 0.038 & 0.036&
 Gemini North  \\ 2020 02 24.601986 & 14 14 31.235 & +27 29 19.63 & 22.9
 & 568 & 0.053 & 0.054& Gemini North  \\ 2020 02 24.608418 & 14 14
 28.325 & +27 29 34.13 & 22.8 & 568 & 0.034 & 0.033& Gemini North  \\
 2020 02 24.621399 & 14 14 22.582 & +27 30 00.12 & 22.4 & 568 & 0.107 &
 0.041& Gemini North  \\ 2020 02 25.148888 & 14 16 23.887 & +27 46 14.71
 & 23.6 & Z23 & 0.2   & 0.2  & NOT              \\ 2020 02 25.156760 &
 14 16 21.045 & +27 46 42.49 & 22.7 & Z23 & 0.08  & 0.08 & NOT          
    \\ 2020 02 25.161406 & 14 16 19.302 & +27 46 58.27 & 22.7 & Z23 &
 0.11  & 0.11 & NOT              \\ 2020 02 26.608875 & 14 18 35.299 &
 +28 40 16.90 & 23.0 & 568 & 0.047 & 0.041& UH 2.2 m          \\ 2020 02
 26.613020 & 14 18 33.500 & +28 40 23.38 & 23.0 & 568 & 0.060 & 0.049&
 UH 2.2 m          \\ 2020 02 26.616949 & 14 18 31.802 & +28 40 29.28 &
 23.2 & 568 & 0.075 & 0.078& UH 2.2 m          \\ 2020 02 26.620998 & 14
 18 30.056 & +28 40 34.90 & 23.2 & 568 & 0.057 & 0.049& UH 2.2 m        
  \\ 2020 02 26.625629 & 14 18 28.070 & +28 40 40.98 & 23.0 & 568 &
 0.055 & 0.045& UH 2.2 m          \\ 2020 02 26.620384 & 14 18 30.334 &
 +28 40 34.01 & 22.76& 568 & 0.05  & 0.05 & CFHT          \\ 2020 02
 26.621564 & 14 18 29.827 & +28 40 35.61 & 22.77& 568 & 0.05  & 0.05 &
 CFHT          \\ 2020 02 26.622763 & 14 18 29.317 & +28 40 37.19 &
 22.74& 568 & 0.05  & 0.05 & CFHT          \\ 2020 02 28.594550 & 14 21
 41.875 & +29 36 30.86 & 22.77& 568 & 0.05  & 0.05 & CFHT          \\
 2020 02 28.595729 & 14 21 41.374 & +29 36 32.58 & 23.19& 568 & 0.05  &
 0.05 & CFHT          \\ 2020 02 28.598103 & 14 21 40.361 & +29 36 36.06
 & 23.23& 568 & 0.05  & 0.05 & CFHT          \\ 2020 03 01.477536 & 14
 23 44.902 & +30 15 25.54 & 22.6 & G37 & 0.256 & 0.268& LDT             
 \\ 2020 03 01.479237 & 14 23 44.291 & +30 15 27.50 & 23.0 & G37 & 0.313
 & 0.283& LDT              \\ 2020 03 01.480279 & 14 23 43.932 & +30 15
 28.52 & 22.8 & G37 & 0.213 & 0.262& LDT              \\ 2020 03
 02.162378 & 14 24 45.080 & +30 30 46.94 & 23.2 & Z23 & 0.07  & 0.07 &
 NOT              \\ 2020 03 02.179266 & 14 24 38.827 & +30 31 16.90 &
 23.1 & Z23 & 0.09  & 0.09 & NOT              \\ 2020 03 02.655003 & 14
 24 20.964 & +30 42 13.62 & 23.32& 568 & 0.05  & 0.05 & CFHT          \\
 2020 03 02.656878 & 14 24 20.269 & +30 42 13.26 & 23.08& 568 & 0.05  &
 0.05 & CFHT          \\ 2020 03 02.658754 & 14 24 19.596 & +30 42 12.91
 & 23.13& 568 & 0.05  & 0.05 & CFHT          \\ 2020 03 04.635008 & 14
 25 48.907 & +31 15 21.90 & 23.30& 568 & 0.05  & 0.05 & CFHT          \\
 2020 03 04.636888 & 14 25 48.213 & +31 15 21.78 & 23.28& 568 & 0.05  &
 0.05 & CFHT          \\ 2020 03 04.638761 & 14 25 47.512 & +31 15 21.58
 & 23.28& 568 & 0.05  & 0.05 & CFHT          \\ 2020 03 05.489286 & 14
 26 26.653 & +31 23 51.85 & 23.1 & G37 & 0.06  & 0.06 & LDT             
 \\ 2020 03 05.490947 & 14 26 26.093 & +31 23 52.37 & 22.9 & G37 & 0.06 
 & 0.06 & LDT              \\ 2020 03 05.492737 & 14 26 25.495 & +31 23
 52.87 & 22.9 & G37 & 0.05  & 0.05 & LDT              \\ 2020 03
 06.240152 & 14 26 41.654 & +31 35 09.84 & 23.3 & Z23 & 0.06  & 0.06 &
 NOT              \\ 2020 03 06.246311 & 14 26 39.515 & +31 35 09.81 &
 23.7 & Z23 & 0.07  & 0.07 & NOT              \\ 2020 03 06.252078 & 14
 26 37.524 & +31 35 09.36 & 23.6 & Z23 & 0.06  & 0.06 & NOT             
 \\ 2020 03 21.462672 & 14 22 58.469 & +33 15 41.15 & 23.77& 568 & 0.05 
 & 0.05 & CFHT          \\ 2020 03 21.464548 & 14 22 57.785 & +33 15
 43.12 & 23.72& 568 & 0.05  & 0.05 & CFHT          \\ 2020 03 21.466429
 & 14 22 57.087 & +33 15 45.00 & 23.59& 568 & 0.05  & 0.05 & CFHT       
   \\ 2020 03 25.199063 & 14 17 34.589 & +33 09 50.72 & 24.0 & Z23 &
 0.12  & 0.12 & NOT              \\ 2020 03 25.210295 & 14 17 30.563 &
 +33 09 39.14 & 24.2 & Z23 & 0.26  & 0.26 & NOT              \\ 2020 03
 25.219775 & 14 17 27.296 & +33 09 28.11 & 23.8 & Z23 & 0.10  & 0.10 &
 NOT              \\ 2020 03 29.161561 & 14 11 40.795 & +32 46 19.67 &
 23.9 & Z23 & 0.08  & 0.05 & NOT              \\ 2020 03 29.171345 & 14
 11 37.179 & +32 46 09.34 & 24.1 & Z23 & 0.09  & 0.06 & NOT             
 \\ 2020 03 29.182286 & 14 11 33.175 & +32 45 56.44 & 24.0 & Z23 & 0.09 
 & 0.06 & NOT              \\ 2020 03 30.376555 & 14 09 54.750 & +32 33
 45.95 & 23.4 & G37 & 0.09  & 0.08 & LDT              \\ 2020 03
 30.379675 & 14 09 53.606 & +32 33 43.93 & 23.5 & G37 & 0.17  & 0.17 &
 LDT              \\ 2020 04 17.072836 & 13 41 35.761 & +26 34 34.21 &
 24.1 & Z23 & 0.089 & 0.102& NOT              \\ 2020 04 17.083259 & 13
 41 32.407 & +26 34 13.83 & 24.3 & Z23 & 0.070 & 0.064& NOT             
 \\ 2020 04 17.093646 & 13 41 29.074 & +26 33 52.11 & 23.8 & Z23 & 0.207
 & 0.207& NOT              \\ 2020 04 29.105118 & 13 25 42.426 & +19 33
 24.57 &      & Z23 & 0.25  & 0.25 & NOT              \\ 2020 04
 29.113261 & 13 25 40.749 & +19 33 00.44 & 24.3 & Z23 & 0.25  & 0.25 &
 NOT              \\ 2020 05 15.936339 & 13 21 25.566 & +07 39 09.07 &
 25.0 & Z23 & 0.15  & 0.15 & NOT              \\ 2020 05 17.244907 & 13
 21 56.056 & +06 44 12.63 & 24.5 & 568 & 0.197 & 0.152 & CFHT           
 \\ 2020 05 17.247704 & 13 21 55.726 & +06 44 05.99 & 24.3 & 568 & 0.281
 & 0.272 & CFHT            \\ 2020 05 20.950021 & 13 23 11.161 & +04 06
 13.56 & 25.6 & Z23 & 0.10  & 0.10 & NOT              \\ 2020 05
 20.967845 & 13 23 09.104 & +04 05 28.13 & 25.4 & Z23 & 0.10  & 0.10 &
 NOT              \\ 2020 05 20.985665 & 13 23 07.102 & +04 04 42.33 &
 25.4 & Z23 & 0.10  & 0.10 & NOT

%    \end{tabular}
    \label{tab:astrometry} \end{longtable}

\begin{longtable}[h]{|l|l|l|l|l|} \caption{Gemini North photometry of
\cdthree{}. The columns are, from left to right: sequential number of
observation;  filter; derived magnitude; instrumental error; zero-point
error.} \\
    Obs. id & Filter & Mag. & $\sigma_{INST}$ & $\sigma_{ZP}$ \\ \hline
    \endfirsthead \multicolumn{4}{r}{\textit{Appendix Table
    \ref{tab:photometry} continued}} \\ Obs. id & Filter & Mag. &
    $\sigma_{INST}$ & $\sigma_{ZP}$ \\ \hline \endhead
    \multicolumn{4}{r}{\textit{Appendix Table \ref{tab:photometry}
    continued on next page}} \\ \endfoot \endlastfoot 1       & $r'$   &
    22.399 & 0.037     & 0.048     \\ 2       & $i'$   & 22.269 & 0.039 
       & 0.054     \\ 3       & $r'$   & 22.413 & 0.035     & 0.048    
    \\ 4       & $g'$   & 23.111 & 0.049     & 0.033     \\ 5       &
    $i'$   & 22.267 & 0.050     & 0.054     \\ 6       & $r'$   & 22.389
    & 0.040     & 0.048     \\ 7       & $g'$   & 23.265 & 0.056     &
    0.033     \\ 8       & $i'$   & 22.340 & 0.053     & 0.054     
   % \end{tabular}
    \label{tab:photometry} \end{longtable}

\begin{longtable}[h]{|l|l|l|l|} \caption{Lightcurve photometric data for
\cdthree{} obtained with the LDT translated to the $r_{\rm P1}$ filter.
The columns are, from left to right: observation date; measured
magnitude; zero-point error; instrumental error.} \\ Date (MJD)        &
Mag($r_{\rm P1}$)     &  $\sigma_{ZP}$    & $\sigma_{INST}$   \\ \hline
\endfirsthead \multicolumn{4}{r}{\textit{Appendix Table
\ref{tab:lightcurve} continued}} \\ Date (MJD)        & Mag($r_{\rm
P1}$)     &  $\sigma_{ZP}$    & $\sigma_{INST}$   \\ \hline \endhead
\multicolumn{4}{r}{\textit{Appendix Table \ref{tab:lightcurve} continued
on next page}} \\ \endfoot \endlastfoot 58906.4881610 & 22.7116 & 0.0228
& 0.1087 \\ 58906.4889677 & 22.6461 & 0.0215 & 0.1050 \\ 58906.4894715 &
22.8355 & 0.0224 & 0.1289 \\ 58906.4899752 & 22.6235 & 0.0227 & 0.1126
\\ 58906.4905406 & 22.9247 & 0.0216 & 0.1407 \\ 58906.4910444 & 22.8095
& 0.0227 & 0.1270 \\ 58906.4915481 & 23.0082 & 0.0226 & 0.1508 \\
58906.4923176 & 22.8036 & 0.0218 & 0.1241 \\ 58906.4928214 & 22.9557 &
0.0227 & 0.1618 \\ 58906.4933251 & 22.6341 & 0.0213 & 0.1121 \\
58906.4938292 & 23.1323 & 0.0232 & 0.1768 \\ 58906.4943329 & 22.8583 &
0.0225 & 0.1346 \\ 58906.4953404 & 22.7895 & 0.0231 & 0.1333 \\
58906.4963479 & 22.8472 & 0.0229 & 0.1373 \\ 58906.4968516 & 23.1161 &
0.0243 & 0.1807 \\ 58906.4973554 & 23.1744 & 0.0229 & 0.1949 \\
58906.5013856 & 22.9624 & 0.0229 & 0.1503 \\ 58906.5018894 & 22.9546 &
0.0232 & 0.1508 \\ 58906.5029427 & 22.8884 & 0.0239 & 0.1488 \\
58906.5034464 & 22.6851 & 0.0234 & 0.1236 \\ 58906.5039502 & 22.9444 &
0.0226 & 0.1473 \\ 58906.5049578 & 23.1792 & 0.0225 & 0.1871 \\
58906.5054615 & 23.0686 & 0.0220 & 0.1513 \\ 58906.5059652 & 22.9580 &
0.0224 & 0.1486 \\ 58906.5074792 & 22.7658 & 0.0227 & 0.1207 \\
58906.5079829 & 22.9301 & 0.0223 & 0.1403 \\ 58906.5089904 & 22.9475 &
0.0224 & 0.1535 \\ 58906.5094942 & 22.6765 & 0.0228 & 0.1160 \\
58906.5099979 & 22.8687 & 0.0225 & 0.1325 \\ 58906.5105017 & 23.0201 &
0.0225 & 0.1490 \\ 58906.5110060 & 22.6124 & 0.0221 & 0.1041 \\
58906.5120135 & 22.6595 & 0.0231 & 0.1040 \\ 58906.5126895 & 23.1251 &
0.0223 & 0.1675 \\ 58906.5131933 & 22.6779 & 0.0220 & 0.1112 \\
58906.5136970 & 22.9829 & 0.0231 & 0.1546 \\ 58906.5142007 & 22.6173 &
0.0221 & 0.1030 \\ 58906.5147052 & 22.9908 & 0.0232 & 0.1551 \\
58906.5152090 & 22.6400 & 0.0225 & 0.1069 \\ 58906.5157131 & 23.0546 &
0.0228 & 0.1527 \\ 58906.5162168 & 22.6199 & 0.0219 & 0.1141 \\
58906.5167207 & 22.6704 & 0.0226 & 0.1151 \\ 58906.5172244 & 23.1026 &
0.0226 & 0.1690 \\ 58906.5177281 & 22.6366 & 0.0217 & 0.1102 \\
58906.5182319 & 22.8375 & 0.0225 & 0.1326 \\ 58906.5187356 & 22.8540 &
0.0232 & 0.1375 \\ 58906.5192395 & 22.8399 & 0.0220 & 0.1317 \\
58906.5197432 & 22.6965 & 0.0221 & 0.1144 \\ 58906.5202471 & 23.0981 &
0.0220 & 0.1653 \\ 58906.5207509 & 22.7457 & 0.0214 & 0.1193 \\
58906.5212546 & 22.8318 & 0.0218 & 0.1292 \\ 58906.5217587 & 23.0903 &
0.0221 & 0.1685 \\ 58906.5222624 & 22.7996 & 0.0216 & 0.1359 \\
58906.5228400 & 22.6896 & 0.0227 & 0.1138 \\ 58906.5233449 & 22.9262 &
0.0230 & 0.1470 \\ 58906.5238487 & 23.1095 & 0.0217 & 0.1704 \\
58906.5243527 & 22.7774 & 0.0224 & 0.1202 \\ 58906.5248564 & 22.7482 &
0.0228 & 0.1234 \\ 58906.5253602 & 22.7669 & 0.0222 & 0.1259 \\
58906.5258639 & 22.9694 & 0.0228 & 0.1472 \\ 58906.5263677 & 22.7342 &
0.0225 & 0.1164 \\ 58906.5283826 & 23.1254 & 0.0226 & 0.1831 \\
58906.5288865 & 22.8278 & 0.0224 & 0.1341 \\ 58906.5293902 & 22.7815 &
0.0221 & 0.1421 \\ 58906.5298940 & 22.7310 & 0.0239 & 0.1272 \\
58906.5319088 & 22.7927 & 0.0230 & 0.1459 \label{tab:lightcurve}
\end{longtable}

\begin{table}[h] \centering \caption{Fits of the $H,G_1,G_2$ system to
the photometric phase curve of \cdthree{}. The first five fits assume
values for the $G_1$ and $G_2$ parameters typical for the spectral types
mentioned in the parenthesis and fit for $H$ only. The last fit allows
for both $H$ and $G_{12}$ parameters to be fitted, but requires
$0<G_{12}<1$, which is a physically meaningful range. The last two
columns provide the weighted root-mean-square (wRMS) value and the
Bayesian Information Criterion with respect to its lowest value
($\Delta$BIC). These results have been computed with the online
calculator available at
http://h152.it.helsinki.fi/HG1G2/.}\label{tab:hg1g2fits}
\begin{tabular}{llllll} Fit type & $H_V$ & $G_1$ & $G_2$ & wRMS &
$\Delta$BIC \\ \hline $H$(E)      & 32.13 & 0.1505 & 0.6005  & 1.632 &
0.000 \\ $H$(S/M)    & 31.79 & 0.2588 & 0.3721  & 1.635 & 0.4107 \\
$H$(P)      & 31.63 & 0.8343 & 0.04887 & 1.658 & 3.310 \\ $H$(C)      &
31.54 & 0.8228 & 0.01938 & 1.661 & 3.686 \\ $H$(D)      & 31.69 & 0.9617
& 0.01645 & 1.661 & 3.763 \\ $H,G_{12}$  & 31.88 & 0.000  & 0.5324  &
1.630 & 4.418 \\ \hline \end{tabular} \end{table}

\begin{figure}[h] \centering
\includegraphics[width=0.6\textwidth]{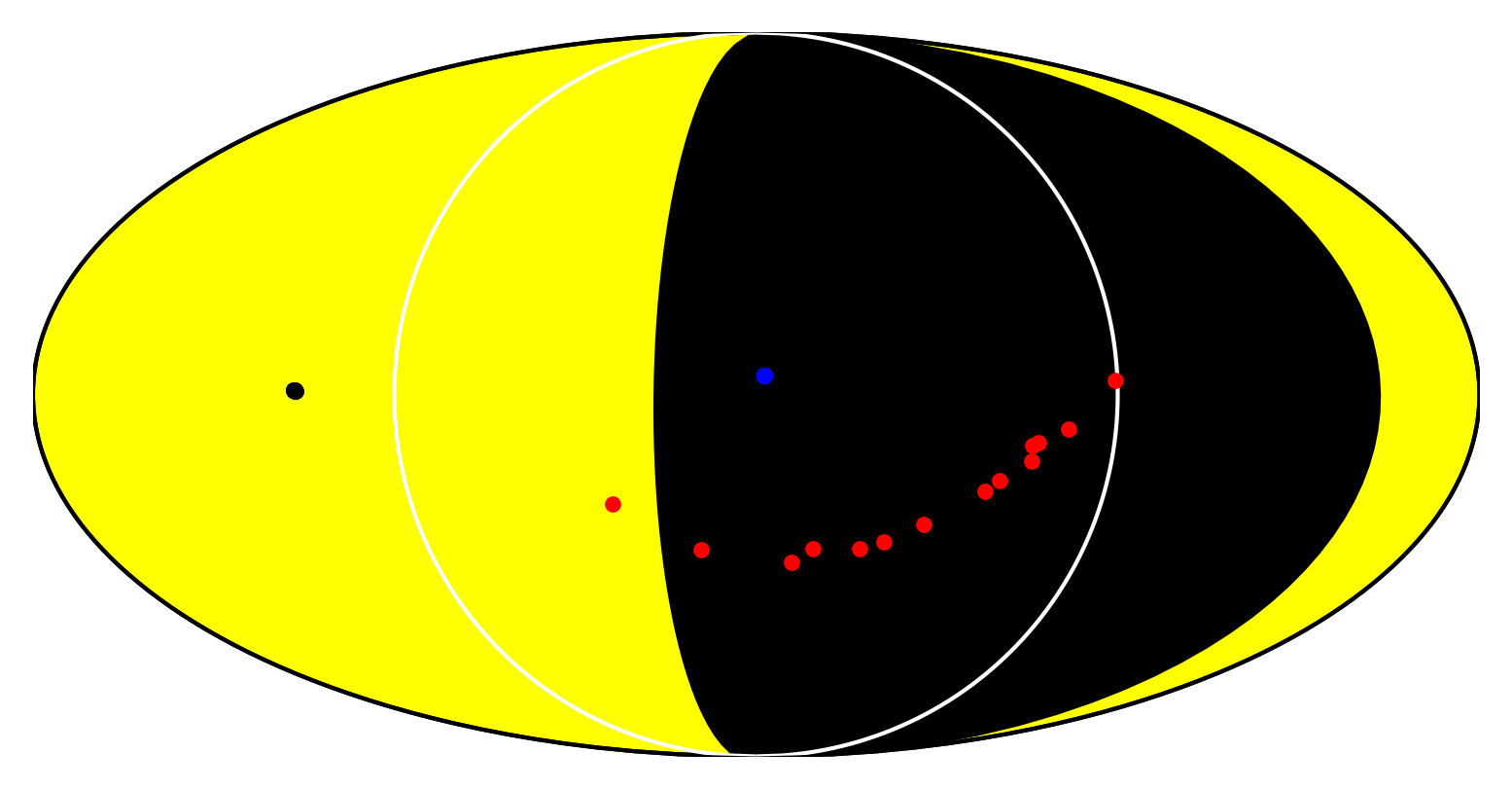}
\caption{Mollweide projection of the surface of the Moon on 2017
September 15. Yellow and black regions show the day and night sides of
the Moon, respectively. Red points show the locations from where the
\cdthree{} samples from the lunar origin hypothesis are ejected. Blue
and black points show the sub-Earth and sub-solar points corresponding
to the sample ejection times. White circle encloses the area of the Moon
visible from Earth.} \label{fig:moonimp} \end{figure}

%\bsp	% typesetting comment
\label{lastpage} \end{document}